# Emergence of Intergranular Tunneling Dominated Negative Magnetoresistance in Helimagnetic Manganese Phosphide Nanorod Thin Films

B. Muchharla[1,*], R. P. Madhogaria[1], D. DeTellem[1], C. M. Hung[1], A. Chanda[1], A. T. Duong[2,3], P. T. Huy[2,3], M. T. Trinh[1], S. Cho[4], S. Witanachchi[1], and M. H. Phan[1,*]

[1] Department of Physics, University of South Florida, Tampa, FL 33620, USA
[2] Faculty of Materials Science and Engineering, Phenikaa University, Hanoi 12116, Vietnam
[3] Phenikaa Research and Technology Institute (PRATI), A&A Green Phoenix Group, 167 Hoang Ngan, Hanoi 13313, Vietnam
[4] Department of Physics and Energy Harvest-Storage Research Center, University of Ulsan, Ulsan 680-749, Republic of Korea

Helical magnets are emerging as a novel class of materials for spintronics and sensor applications; however, research on their charge– and spin–transport properties in a thin film form is less explored. Herein, we report the temperature and magnetic field–dependent charge transport properties of a highly crystalline MnP nanorod thin film over a wide temperature range (2 K < $T$ < 350 K). The MnP nanorod films of ~100 nm thickness were grown on Si substrates at 500 °C using molecular beam epitaxy. The temperature–dependent resistivity $\rho(T)$ data exhibits a metallic behavior (d$\rho$/d$T$ > 0) over the entire measured temperature range. However, large negative magnetoresistance ($\Delta\rho/\rho$) of up to 12 % is observed below ~50 K at which the system enters a stable helical (screw) magnetic state. In this temperature regime, the $\Delta\rho(H)/\rho(0)$ dependence seems to show a magnetic field–manipulated phase coexistence. The observed magnetoresistance is dominantly governed by the intergranular spin dependent tunneling mechanism. These findings pinpoint a correlation between the transport and magnetism in this helimagnetic system.

*Corresponding authors: BaleeswaraiahMuchharla@gmail.com (B.M.); phanm@usf.edu (M.H.P)

1. **Introduction**

   Manganese phosphide (MnP) is a complex magnetic material with multiple exotic magnetic phases. MnP displays various magnetic orderings such as para, ferro, screw and fan structures, depending on the temperature and applied magnetic field. MnP has an orthorhombic crystal structure with lattice parameters $a$ = 5.916 Å, $b$ = 3.173 Å, and $c$ = 5.260 Å. The unit cell contains four Mn atoms and four P atoms. MnP is a 3d transition metal compound that exhibits a wide range of properties including helimagnetism[1–4], thermoelectricity[5], magneto–caloric[6,7], magneto–optical[8] and superconductivity[9–12]. Single crystal MnP undergoes a paramagnetic (PM) to ferromagnetic (FM) phase transition around room temperature and displays another phase transition from the FM to helical (screw) phase around 50 K in the absence of an external magnetic field[6]. The MnP crystal displays magnetic anisotropy with $c$–axis being the easy direction of magnetization, $b$–axis is the intermediate and $a$–axis is the hardest one. The phase transition temperatures of the MnP single crystal can be altered by doping with small amounts of Co. With an amount of 0.3 at.% Co doping, both the transition temperatures are lowered by about 5 K, and further increase in the Co concentration to 5% decreases the PM – FM transition to 240 K while vanishing the helimagnetic (HM) transition[13]. In the same study, the topological Hall effect (THE) was observed in the FAN phase of MnP when the magnetic field was applied along the $b$–axis. Unlike conventional and anomalous Hall effects, THE is typically observed in a phase with the lattice formation of a skyrmion, stabilized by the Dzyaloshinsky – Moriya (DM) interaction, and it appears to occur when conduction electrons acquire the Berry phase while passing through the skyrmion lattice. The THE observed in MnP was reported to decay with 5% doping of Co at the Mn sites[13]. Single crystal MnP, prepared by vapor phase transport, was experimentally investigated to exhibit the de Haas–van Alphen (dHvA) effect [14,15]. The dHvA oscillations were observed with applied fields up to 94 kOe along the $c$–axis at ~1.1 K. The Shubnikov–de Haas like oscillation was also observed in the high purity single crystal of MnP at 1.5 K in applied fields up to 80 kOe[16]. Recently, Jiang *et al.* have reported the control of the spin helicity in MnP by using an electric current and a magnetic field which could be utilized in applications such as magnetic memories based on the spin internal degrees of freedom[17].

While previous studies focused on the transport and spin properties of MnP single crystals[13–18], device applications would be realized mostly on thin films[19]. We have recently reported a comprehensive magnetic study on highly crystalline MnP nanorod films in which the magnetic phase diagrams have been constructed demonstrating the co–existing magnetic phases for the in–plane and out–of–plane field orientations [20]. These findings have motivated us to investigate the charge and spin transport properties of the MnP nanorod film over a wide temperature range (2 K $<T<$ 350 K) in magnetic fields up to 8 T. In agreement with the previously reported transport results on MnP single crystals[13,17,19,21], a metallic conducting nature is also observed for the MnP nanorod film. However, we have observed two distinct temperature regimes, above and below the FM–HM transition temperature (~50 K), where electrical resistivity ($\rho$) shows a linear temperature dependence for $T>$ 50 K and a non–linear power law dependence for $T<$ 50 K. Interestingly, we have observed, for the first time, a large negative magnetoresistance ($\Delta\rho/\rho$) of up to 12 % and a magnetic field–manipulated CONE + FAN phase coexistence in this low temperature region ($T<$ 50 K). These findings relate the charge–spin properties characteristic for *itinerant* helimagnetic systems like MnP.

2. **Experimental**

MnP nanorod films were grown on Si (*100*) substrates using molecular beam epitaxy while maintaining the substrate at a temperature of 500 °C. The details of MnP nanorod thin film growth can be found elsewhere[19]. The crystal phase characterization of the MnP thin films was carried out using Bruker AXS powder X–ray diffractometer (XRD) with Cu–K$_\alpha$ radiation at room temperature. Field emission scanning electron microscopy (FE–SEM) was performed to observe microstructural variation in the films grown at different temperatures. Temperature dependent transport measurements were carried out using four-probe measurement system consisting of Keithley 2400 source meter and Keithley 2182a nanovoltmeter integrated with the Physical Property Measurement System (PPMS) from Quantum Design. Temperature– and magnetic field–dependent electrical resistivity and current–voltage (*I*–*V*) characteristic curves were measured within the wide range of temperature (2 K < *T* < 350 K) and applied magnetic field (0 T < $\mu_0H$ < 8 T).

**3. Results and discussion**

The XRD pattern of the MnP thin film sample displays the orthorhombic MnP peaks as defined in Fig. 1(a). The energy dispersive X-ray spectroscopy (EDS) spectrum of the MnP film, as shown in an inset of Fig. 1(a), consistently confirms the presence of Mn and P elements in the MnP film grown on the Si substrate. Surface morphology of the MnP film was characterized by FE–SEM, as presented in Fig. 1(b), indicating closely packed MnP nanorods grown vertically out of the plane on the Si substrate. The inset of Fig. 1(b) displays an AFM image of the MnP film. A typical four-probe method was utilized as depicted in the schematic (Fig. 2(a)) to acquire the transport data of the MnP film device. In order to understand the transport nature of the MnP film, we measured the temperature dependent resistivity over a wide range (10 K < *T* < 305 K) and the result is displayed in Fig. 2(b). The room temperature resistivity is of 4.25 x $10^{-4}$ Ωcm, and the value decreases with lowering temperature (Fig. 2(b)) indeed shows a metallic behavior for this film in the entirely measured temperature range. This type of metallic nature is in agreement with previous reports on MnP[13,17,19,21]. In the absence of an external magnetic field, bulk MnP undergoes a PM–FM phase transition at ~290 K and then stabilizes into a stable helical (screw) phase below ~50 K. The observed *ρ*(*T*) behavior of the MnP film indicates that there exist two slope changes right around these magnetic transitions, indicating a correlation between the conductivity and magnetism in this system. These phase transitions are clearly evidenced with the slope changes on the derivative of resistivity versus temperature plot displayed (blue circles) on the right y-axis of Fig. 2(b). Temperature–dependent magnetization ($M - T$) behavior is displayed in Fig. 2(d) under field–cooled warming protocol for 2 K < $T$ < 370 K. A PM – FM phase transition is clearly observed around 300 K, followed by the FM transition to the stable helical (screw) state transition at ~50 K as shown in Fig. 2(d). Note that the onset temperature of the FM to HM transition occurs at ~100 K and stabilizes to the helical phase below 50 K. The derivative of zero-field-cooled magnetization with respect to temperature plotted (blue circles) on the right y-axis of Fig. 2(d) represents the PM-FM phase transition around 300 K, with the FM to HM onset at ~100 K and stabilized to the HM phase below 50 K. Previous studies have shown similar magnetic phase transitions in helimagnetic MnP samples. For instance, 2D MnP single crystals grown using a conventional vapor deposition (CVD) technique exhibited a paramagnetic-ferromagnetic transition at 303 K and the helical magnetic (screw) state at 38 K [22]. Whereas Andrés *et al.* have reported a large variation in the ferromagnetic-screw transition temperature for bulk MnP (~47 K), the MnP thin film (~67 K), and the MnP nanocrystals (~82 K) embedded in the GaP epilayers [23]. Note that the phase transition temperatures varied, depending on the size and shape of the nanocrystals present in the samples. In the present study, it is the formation of MnP nanorods that led to the occurrence of the onset of the screw phase transition at ~100 K.

In this study, the temperature dependence of electrical resistivity is broken down into two distinct temperature regimes, separated at ~50 K, below which the system enters a stable helical (screw) magnetic state. For $T > $ ~50 K, a monotonic decrease in electrical resistivity is seen with decrease in temperature. For $T < 50$ K, however, a non–linear temperature dependence of electrical resistivity with decrease in the temperature is observed. Electrical resistivity of a metallic ferromagnet[24] can be explicitly described as

$$\rho(T) = \rho_0 + \rho_{e-e} + \rho_{e-ph} + \rho_{e-mag}, \qquad (1)$$

where $\rho_0$ is a temperature independent term which corresponds to the contribution from impurity scattering, $\rho_{e-e}$ accounts for the Coulombic interaction between the conduction electrons ($T^2$ dependence), $\rho_{e-ph}$ is associated with the electron-phonon interaction, and $\rho_{e-mag}$ accounts for electron-magnon interactions. In case of two-magnon scattering mechanism, $\rho_{e-mag}$ varies as $\sim T^{4.5}$ at low temperatures and $\sim T^{3.5}$ in the high temperature region[25]. It should be recalled that bulk MnP is a helimagnet, which exhibits complex magnetic phases with respect to temperature and magnetic field. In the absence of an external magnetic field, bulk MnP undergoes a PM–FM phase transition at ~290 K and then stabilizes into a helical (screw) phase below ~50 K [20]. It is clear from the Fig. 2(c) that $\rho(T)$ for our MnP film undergoes two consecutive slope changes: one around the PM-FM phase transition (~290 K) and another around ~50 K that coincides with the FM to screw phase transformation. Therefore, the electrical transport mechanism in our MnP film is strongly correlated to the nature of the magnetic ground state. Distinct magnon dispersion is expected above and below 50 K, which also indicates the different nature of the electron-magnon scattering mechanisms in these two different temperature regimes. Considering that the electron-phonon interaction has significant contribution only in the high temperature region, we fitted the $\rho(T)$ curve for 50 K < $T \leq$ 300 K (solid red line) with the following expression:

$$\rho(T) = \rho_0 + \rho_{e-e} + \rho_{e-ph} + \rho_{e-mag} = \rho_0 + \rho_1 T + \rho_2 T^2 + \rho_{3.5} T^{3.5} \qquad (2)$$

From the fit, we obtained $\rho_0$ = 1.32 x 10$^{-5}$ Ω cm, $\rho_1$ = 9.72 x 10$^{-7}$ Ω cm. K$^{-1}$, $\rho_2$ = 3.74 x 10$^{-9}$ Ω cm. K$^{-2}$ and $\rho_{3.5}$ = 4.63 x 10$^{-13}$ Ω cm. K$^{-3.5}$. Ignoring the contribution of electron-phonon interaction at low temperatures, we fitted the $\rho(T)$ curve in the temperature regime 10 K $\leq T \leq$ 50 K (black line) with the following expression,

$$\rho(T) = \rho_0 + \rho_{e-e} + \rho_{e-mag} = \rho_0 + \rho_2 T^2 + \rho_{4.5} T^{4.5} \qquad (3)$$

The values of the coefficients $\rho_0$, $\rho_2$ and $\rho_{4.5}$ obtained from the fit are 5.55 x $10^{-5}$ $\Omega$ cm, 5.63 x $10^{-9}$ $\Omega$ cm. $K^{-2}$ and 9.35 x $10^{-14}$ $\Omega$ cm. $K^{-4.5}$, respectively. It is evident that the electron-electron scattering is the dominating mechanism, as compared to the electron-magnon scattering, governing the electrical transport in the MnP film. It should be recalled that bulk MnP is a helimagnet, which exhibits complex magnetic phases with respect to temperature and magnetic field[17]. Transport studies on needle–shaped MnP single crystals grown using Sn–flux method by Cheng *et al.* also showed a $T^2$ dependence of electrical resistivity at low temperatures and interestingly a superconducting behavior upon the application of a high pressure (~ 8 GPa) at ~1 K, the MnP system was driven into the superconducting state[11].

Magneto–transport measurements were carried out on the same sample by applying magnetic field perpendicular to the plane of the film. Temperature dependence of electrical resistivity, $\rho(T)$ was studied in the temperature range 10 K $\leq T \leq$ 305 K at different magnetic field strengths as shown in the main panel of Fig. 3(a). The trend of $\rho(T)$ in presence of the external magnetic field is similar to that observed in zero magnetic field. However, the value of $\rho$ drops significantly with increasing magnetic field strength at low temperatures especially below ~50 K ($\rho$ drops from 0.05 m$\Omega$ cm for $\mu_0 H = 0$ T to 0.04 m$\Omega$ cm for $\mu_0 H = 5$ T at $T = 10$ K), indicating that the electrical transport is strongly correlated to the HM state at low temperatures. Decrease in electrical resistivity with the application of magnetic field resulted in a negative magnetoresistance. The temperature dependence of magnetoresistance was estimated using the percentage change as, MR(%) = [{$\rho(T, \mu_0 H) - \rho(T, \mu_0 H = 0$ T)}/{$\rho(T, \mu_0 H = 0$ T}] x 100. The inset of Fig. 3(a) shows the temperature dependence of magnetoresistance at $\mu_0 H = 5$ T which clearly shows a drastic enhancement of negative magnetoresistance below 50 K. To verify this enhancement of magnetoresistance, we have also performed the magnetic field dependence of magnetoresistance at different temperatures between $T$

= 300 and 2 K, as shown in Fig. 3(b). The magnetic field dependence of magnetoresistance is estimated using the relative resistivity ratio: MR(%) = [{$\rho(H) - \rho(H = 0)$}/{$\rho(H = 0)$}] x 100. It is evident that the value of negative magnetoresistance is dramatically enhanced below 50 K, which underscores a strong correlation between the transport mechanism and the magnetic phase transition from the FM to the screw phase. The value of MR(%) reaches the highest value of 12.5% at 2 K which is nearly consistent with the temperature-dependent magnetoresistance data presented in the inset of Fig. 3(a). A dominant butterfly shape of magnetoresistance isotherms with considerable hysteresis appears to occur at low temperatures below 50 K. The inset of Fig. 3(b) displays magnetoresistance versus magnetic field at 2 K with magnetic field sweep directions were indicated by red and blue arrows. The hysteresis becomes strongly suppressed for 50 K $< T <$ 100 K and diminished for $T >$ ~100 K at which the onset temperature of the FM to HM transition is observed.

The nature of the magnetic field dependence of magnetoresistance for our MnP nanorod thin films has certain similarities to spin-dependent tunneling magnetoresistance observed in ferromagnetic nanoclusters embedded in nonmagnetic insulating/semiconducting matrix e.g., Co–SiO$_2$ granular films[26], ferromagnetic cobaltite La$_{0.85}$Sr$_{0.15}$CoO$_3$ [27], etc. The probability of tunneling of conduction electrons depends on the relative orientation of the magnetic moments of neighboring grains, which increases when these magnetic moments are aligned parallel to each other. Such a conducting mechanism is quite plausible in our MnP system because of several interfaces formed along the aligned nanorods. In the absence of an external magnetic field, the magnetic moments of the adjacent grains are randomly oriented, thus limiting conduction electrons to tunnel between these grains. As the applied magnetic field increases, these magnetic moments will be aligned with the field direction, giving rise to the enhanced intergranular tunneling of conduction electrons and hence the lower resistance state. In case of inelastic tunneling of electrons across a single potential barrier, the tunneling magnetoresistance can be expressed as[28]

$$\frac{\Delta \rho}{\rho} = -\frac{P^2 m^2}{1 + P^2 m^2} \tag{4}$$

where $P$ is the degree of spin polarization and $m$ is the reduced magnetization. For a granular film, it has been theoretically shown that, $\frac{\Delta \rho}{\rho} \propto m^2$ [28]; which simply indicates that the field dependence of magnetoresistance should follow the trend of squared magnetization. In Fig. 3(c), we plotted the magnetoresistance isotherm at $T = 10$ K together with the $M(H)$ hysteresis loop measured in the out-of-plane (OOP) configuration at the same temperature. It is evident that the peak in magnetoresistance occurs around the coercive field, which is followed by a slope change in magnetoresistance that coincides with the high field switching behavior in the OOP $M(H)$. As shown in Fig. 3(d), the magnetoresistance isotherm exactly follows the field dependence of reduced squared magnetization $(M/M_{6T})^2$ at $T = 10$ K, which indicates that the observed magnetoresistance in our MnP nanorod thin film is dominated by the intergranular spin-dependent tunneling mechanism.

To affirm the negative MR effect and gain further insight into the correlation between the transport and magnetism in the MnP film, $I$–$V$ characteristic curves were also measured by varying temperature (10 – 300 K) and magnetic field (0 – 8 T). All $I$–$V$ curves measured for the MnP sample display linear behavior indicating the ohmic nature of the sample. As can be observed in Fig. 4(a), in the absence of an applied magnetic field, the slopes of the $I$–$V$ curves increase with decrease in the temperature, indicating the increase/decrease in conductivity/resistivity with lowering temperature, which is in full agreement with that shown in Fig. 2(a). A similar behavior of the $I$–$V$ curves taken at different temperatures is observed in the presence of 8 T as shown Fig. 4(b). Figs. 4(c) and (d) display the $I$–$V$ curves at constant temperatures ((c) at 300 K and (d) at 10 K) with magnetic field varying from 0 to 8 T. While at room temperature, the increase in the applied magnetic field, the slopes of the $I$–$V$ curves slightly increased, the change at 10K is more pronounced indicating the decrease in electrical resistivity with magnetic field, as shown in Fig. 4(c–d). That results in the sizable MR effect, in full agreement with that shown in Fig. 3(c).

In order to establish this correlation, resistivity values obtained from the slopes of the *I–V* characteristics were first plotted against temperature (2 – 200 K) and magnetic field (0 – 8 T) in a 2D–surface plot as shown in Fig. 5(a). As expected, the electrical resistivity decreases with decreasing temperature and increasing magnetic field. Fig. 5(c) displays the magnetoresistance data for the same ranges of temperature and magnetic field as in Fig. 5(a). An increase in negative magnetoresistance is also observed with decrease in the temperature and increase in the magnetic field. Figs. 5(b) and (d) display the 2D–surface plots of the resistivity and magnetoresistance data in the temperature range (200 – 350 K) and in the magnetic field range (0 – 8 T). A stable FM phase is observed from ~230 K down to ~100 K which is the onset temperature of the FM – HM transition. Below 100 K, a mixed phase (FM + HM) coexists, and the HM phase is dominant below ~50 K. Very interesting features are observed below 50 K from the magnetoresistance data as shown in Fig. 5(c). A constant negative magnetoresistance of about 2% was observed at these temperatures when the applied magnetic field was less than 1.5 T. Further increase in the applied magnetic field (1.5 T < $\mu_0 H$ < 8 T) predominantly increases the negative magnetoresistance (reaching ~12% at 2 K and 8 T). A distinguishable change in the magnetoresistance behavior at ~50 K is related to the transition from the FM to the (complete) screw phase. At a given temperature (e.g., 2 K), we have observed an obvious change in the MR ratio at the critical fields at which the sample undergoes different magnetic transitions (e.g., CONE, FAN). Interestingly, at ~300 K, a negative to positive magnetoresistance is observed for the MnP sample, which is associated with the PM to FM phase transition that occurs at the same temperature. These findings point to a new possibility of controlling the charge transport in HM systems by controlling an external magnetic field. From a fundamental research perspective, our charge–spin–transport study also suggests it as a useful probe of the competing phases' coexistence and the magnetic field–driven conversion of these phases in HM systems like MnP, adding complementary information to the complex magnetic phase diagrams of these systems.

## 4. Conclusion

In conclusion, the temperature and magnetic field–dependent charge transport properties of the MnP nanorod thin film have been studied systematically. The MnP film exhibits a metallic behavior over the entire measured temperature range (5 – 350 K). The intrinsic electrical resistivity of the MnP film displays the $T^2$ dependence at low temperatures (< 50 K) and a linear dependence at higher temperatures. A large negative magnetoresistance of up to 12% is observed in the helical magnetic regime ($T <$ ~50 K), under the application of high magnetic fields up to 8 T. The magnetic transitions from the PM to FM phase and from the FM to stable HM phase, as well as the field–driven conversion of magnetic phases (e.g., CONE, FAN) are also revealed from the magneto–transport data. It has been established that the low temperature magnetoresistance is dominated by the intergranular spin dependent tunneling mechanism. Our study pinpoints the correlation between the transport and magnetism in this helimagnetic system.

**Acknowledgments**

Research at the University of South Florida was supported by the U.S. Department of Energy, Office of Basic Energy Sciences, Division of Materials Sciences and Engineering under Award No. DE-FG02-07ER46438.

**References**

[1]  G.P. Felcher, Magnetic Structure of MnP, J. Appl. Phys. 37 (1966) 1056–1058. https://doi.org/10.1063/1.1708333.

[2]  H. Obara, Y. Endoh, Y. Ishikawa, T. Komatsubara, Magnetic Phase Transition of MnP Under Magnetic Field, J. Phys. Soc. Japan. 49 (1980) 928–935. https://doi.org/10.1143/JPSJ.49.928.

[3]  M. Matsuda, F. Ye, S.E. Dissanayake, J.-G. Cheng, S. Chi, J. Ma, H.D. Zhou, J.-Q. Yan, S. Kasamatsu, O. Sugino, T. Kato, K. Matsubayashi, T. Okada, Y. Uwatoko, Pressure dependence of the magnetic ground states in MnP, Phys. Rev. B. 93 (2016) 100405. https://doi.org/10.1103/PhysRevB.93.100405.

[4]  E.E. Huber, D.H. Ridgley, Magnetic Properties of a Single Crystal of Manganese Phosphide, Phys. Rev. 135 (1964) A1033–A1040. https://doi.org/10.1103/PhysRev.135.A1033.

[5]  T. Suzuki, Magnetic Field Effects on the Thermoelectric Power in Mnp Single Crystal, J. Phys. Soc. Japan. 26 (1969) 279–283. https://doi.org/10.1143/JPSJ.26.279.

[6]  M.S. Reis, R.M. Rubinger, N.A. Sobolev, M.A. Valente, K. Yamada, K. Sato, Y. Todate, A. Bouravleuv, P.J. von Ranke, S. Gama, Influence of the strong magnetocrystalline anisotropy on the magnetocaloric properties of MnP single crystal, Phys. Rev. B. 77 (2008) 104439. https://doi.org/10.1103/PhysRevB.77.104439.

[7]  R.A. Booth, S.A. Majetich, Crystallographic orientation and the magnetocaloric effect in MnP, J. Appl. Phys. 105 (2009) 07A926. https://doi.org/10.1063/1.3072022.

[8]  G. Monette, N. Nateghi, R.A. Masut, S. Francoeur, D. Ménard, Plasmonic enhancement of the magneto-optical response of MnP nanoclusters embedded in GaP epilayers, Phys. Rev. B. 86 (2012) 245312. https://doi.org/10.1103/PhysRevB.86.245312.

[9]  X. Chong, Y. Jiang, R. Zhou, J. Feng, Pressure dependence of electronic structure and superconductivity of the MnX (X = N, P, As, Sb), Sci. Rep. 6 (2016) 21821.

https://doi.org/10.1038/srep21821.

[10] Y. Wang, Y. Feng, J.-G. Cheng, W. Wu, J.L. Luo, T.F. Rosenbaum, Spiral magnetic order and pressure-induced superconductivity in transition metal compounds, Nat. Commun. 7 (2016) 13037. https://doi.org/10.1038/ncomms13037.

[11] J.-G. Cheng, K. Matsubayashi, W. Wu, J.P. Sun, F.K. Lin, J.L. Luo, Y. Uwatoko, Pressure Induced Superconductivity on the border of Magnetic Order in MnP, Phys. Rev. Lett. 114 (2015) 117001. https://doi.org/10.1103/PhysRevLett.114.117001.

[12] J. Cheng, J. Luo, Pressure-induced superconductivity in CrAs and MnP, J. Phys. Condens. Matter. 29 (2017) 383003. https://doi.org/10.1088/1361-648X/aa7b01.

[13] Y. Shiomi, S. Iguchi, Y. Tokura, Emergence of topological Hall effect from fanlike spin structure as modified by Dzyaloshinsky-Moriya interaction in MnP, Phys. Rev. B. 86 (2012) 180404. https://doi.org/10.1103/PhysRevB.86.180404.

[14] M. Ohbayashi, T. Komatsubara, E. Hirahara, De Haas-van Alphen Effect in Manganese Phosphide, J. Phys. Soc. Japan. 40 (1976) 1088–1094. https://doi.org/10.1143/JPSJ.40.1088.

[15] S. Kawakatsu, M. Kakihana, M. Nakashima, Y. Amako, A. Nakamura, D. Aoki, T. Takeuchi, H. Harima, M. Hedo, T. Nakama, Y. Ōnuki, De Haas–van Alphen Experiment and Fermi Surface Properties in Field-Induced Ferromagnetic State of MnP, J. Phys. Soc. Japan. 88 (2019) 044705. https://doi.org/10.7566/JPSJ.88.044705.

[16] A. Takase, T. Kasuya, High Field Magnetoresistance in MnP, J. Phys. Soc. Japan. 49 (1980) 489–492. https://doi.org/10.1143/JPSJ.49.489.

[17] N. Jiang, Y. Nii, H. Arisawa, E. Saitoh, Y. Onose, Electric current control of spin helicity in an itinerant helimagnet, Nat. Commun. 11 (2020) 1601. https://doi.org/10.1038/s41467-020-15380-z.

[18] Y. Matsumura, E. Narita, E. Hirahara, Effect of Uniaxial Pressure on the Magnetic Transition

Temperatures in Manganese Phosphide, J. Phys. Soc. Japan. 38 (1975) 1264–1269. https://doi.org/10.1143/JPSJ.38.1264.

[19] A.-T. Duong, T.M.H. Nguyen, D.-L. Nguyen, R. Das, H.-T. Nguyen, B.T. Phan, S. Cho, Enhanced magneto-transport and thermoelectric properties of MnP nanorod thin films grown on Si (1 0 0), J. Magn. Magn. Mater. 482 (2019) 287–291. https://doi.org/10.1016/j.jmmm.2019.03.072.

[20] R.P. Madhogaria, C.-M. Hung, B. Muchharla, A.T. Duong, R. Das, P.T. Huy, S. Cho, S. Witanachchi, H. Srikanth, M.-H. Phan, Strain-modulated helimagnetism and emergent magnetic phase diagrams in highly crystalline MnP nanorod films, Phys. Rev. B. 103 (2021) 184423. https://doi.org/10.1103/PhysRevB.103.184423.

[21] P. Zheng, Y.J. Xu, W. Wu, G. Xu, J.L. Lv, F.K. Lin, P. Wang, Y. Yang, J.L. Luo, Orbital-dependent charge dynamics in MnP revealed by optical study, Sci. Rep. 7 (2017) 14178. https://doi.org/10.1038/s41598-017-14648-7.

[22] X. Sun, S. Zhao, A. Bachmatiuk, M.H. Rümmeli, S. Gorantla, M. Zeng, L. Fu, 2D Intrinsic Ferromagnetic MnP Single Crystals, Small. 16 (2020) 2001484. https://doi.org/10.1002/smll.202001484.

[23] A. de Andrés, A. Espinosa, C. Prieto, M. García-Hernández, R. Ramírez-Jiménez, S. Lambert-Milot, R.A. Masut, MnP films and MnP nanocrystals embedded in GaP epilayers grown on GaP(001): Magnetic properties and local bonding structure, J. Appl. Phys. 109 (2011) 113910. https://doi.org/10.1063/1.3580270.

[24] B. Venkateswarlu, P.V. Midhunlal, P.D. Babu, N.H. Kumar, Magnetic and anomalous electronic transport properties of the quaternary Heusler alloys Co2Ti1−Fe Ge, J. Magn. Magn. Mater. 407 (2016) 142–147. https://doi.org/10.1016/j.jmmm.2016.01.059.

[25] K. Kubo, N. Ohata, A Quantum Theory of Double Exchange. I, J. Phys. Soc. Japan. 33

(1972) 21–32. https://doi.org/10.1143/JPSJ.33.21.

[26]  S. Sankar, A.E. Berkowitz, D.J. Smith, Spin-dependent transport of Co−SiO2 granular films approaching percolation, Phys. Rev. B. 62 (2000) 14273–14278. https://doi.org/10.1103/PhysRevB.62.14273.

[27]  J. Wu, J.W. Lynn, C.J. Glinka, J. Burley, H. Zheng, J.F. Mitchell, C. Leighton, Intergranular Giant Magnetoresistance in a Spontaneously Phase Separated Perovskite Oxide, Phys. Rev. Lett. 94 (2005) 037201. https://doi.org/10.1103/PhysRevLett.94.037201.

[28]  J. Inoue, S. Maekawa, Theory of tunneling magnetoresistance in granular magnetic films, Phys. Rev. B. 53 (1996) R11927–R11929. https://doi.org/10.1103/PhysRevB.53.R11927.

**Figure Captions**

**Fig. 1** Structural and morphological characterization of the MnP film: (a) XRD pattern of the MnP sample. The inset displays EDS spectra of the MnP film.; (b) SEM image of closely packed MnP nanorods grown vertically on the Si substrate. The inset displays an AFM image of the MnP film. The scale bar on the AFM image is 500 nm.

**Fig. 2** Temperature–dependent electrical transport properties of the MnP film: (a) Typical device schematic indicating the device structure with four-probe configuration; (b) Temperature dependence of the electrical resistivity in the temperature range 10 K < $T$ < 305 K. The plot on the right y-axis (blue circles) displays the derivative of resistivity versus temperature; (c) Temperature dependence of the electrical resistivity plot indicating the screw and FM temperature regions; (d) Temperature dependence of magnetization, $M$ vs. $T$ curves, under a zero field–cooled–warming protocol. The plot on the right y-axis (blue circles) displays the derivative of zero field cooled magnetization with respect to temperature.

**Fig. 3** Temperature and magnetic field dependent magneto-transport properties of the MnP film; (a) Temperature dependence of the resistivity in the temperature range 10 K < $T$ < 305 K in the presence of discrete magnetic fields. The inset shows the temperature dependence of magnetoresistance at $\mu_0 H$ = 5 T; (b) magnetoresistance versus magnetic field at various temperatures. The inset displays magnetoresistance versus magnetic field at 2 K. Red and blue arrows indicate the direction of magnetic field sweep; (c) comparison of magnetoresistance (left-y-scale) and out-of-plane (OOP) magnetization (right-y-scale) at $T$ = 10 K and (d) correlation between magnetoresistance (left-y-scale) and reduced squared magnetization (right-y-scale) at $T$ = 10 K signifying a dominant role of intergranular tunneling.

**Fig. 4** $I$–$V$ characteristics of the MnP film: at (a) $\mu_0 H$ = 0 T and (b) $\mu_0 H$ = 8 T with varying temperature from 10 to 300 K; (c) at $T$ = 300 K and (d) $T$ = 10 K with varying magnetic field from 0 to 8 T.

**Fig. 5** Magnetic field and temperature dependent 2D–surface plots of (a) resistivity and (c) magnetoresistance in the temperature range 2 K $< T <$ 200 K and (b) resistivity and (d) magnetoresistance in the temperature range 200 K $< T <$ 350 K.

**Figure 1**

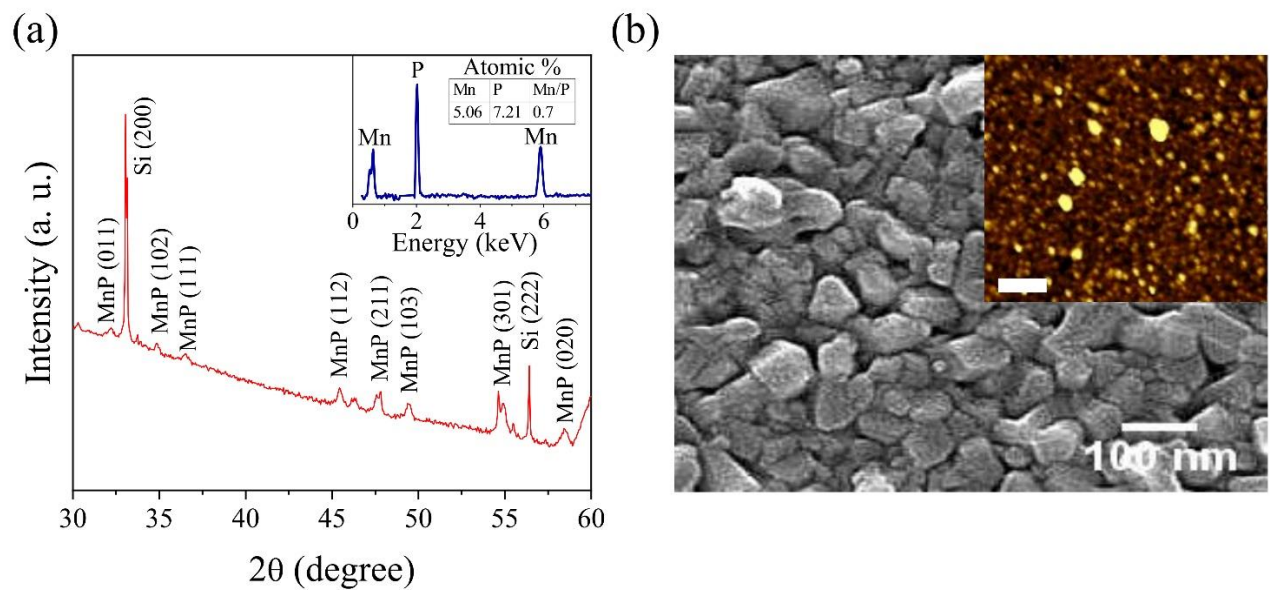

**Figure 2**

(a) 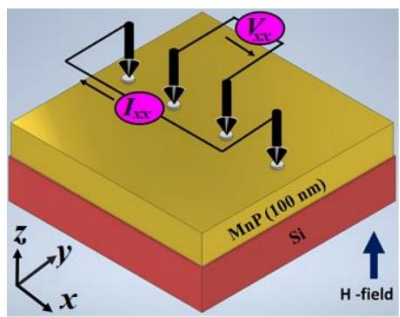

(b) 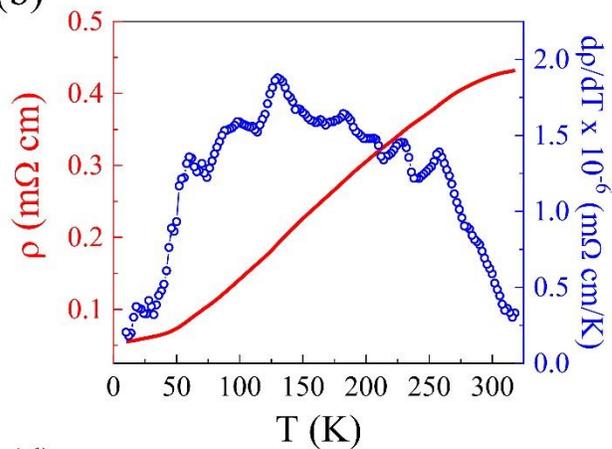

(c) 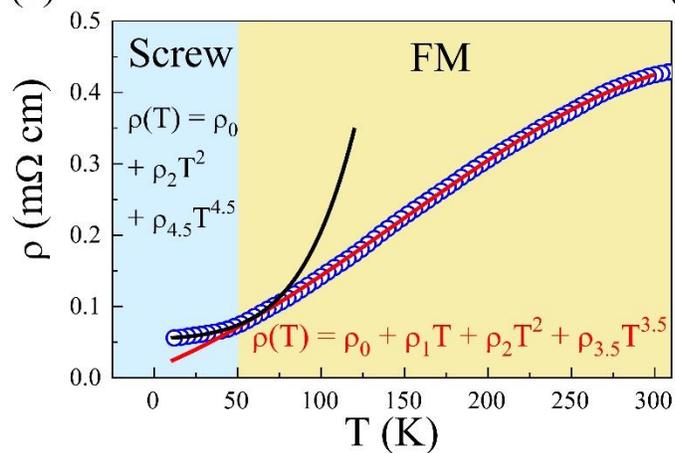

(d) 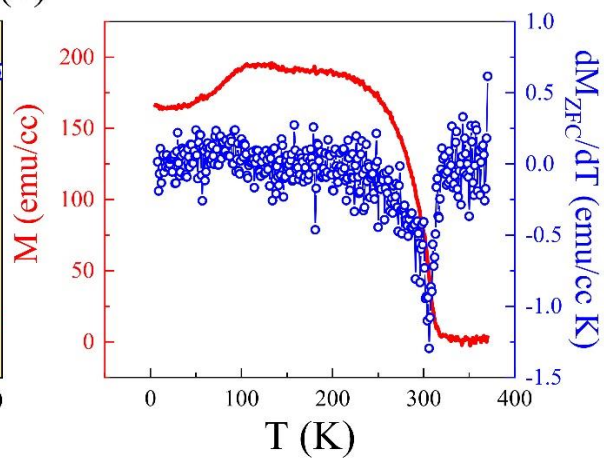

**Figure 3**

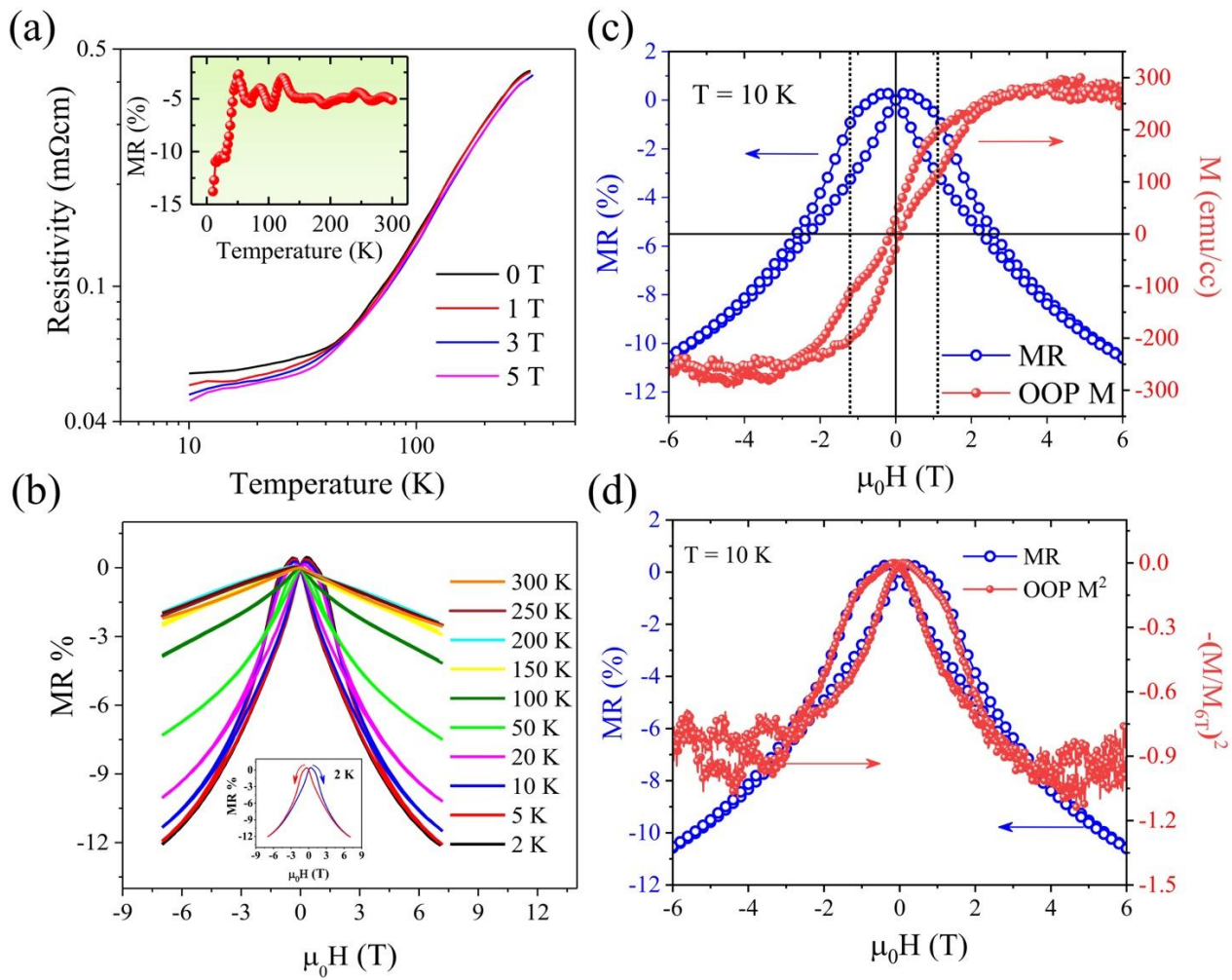

**Figure 4**

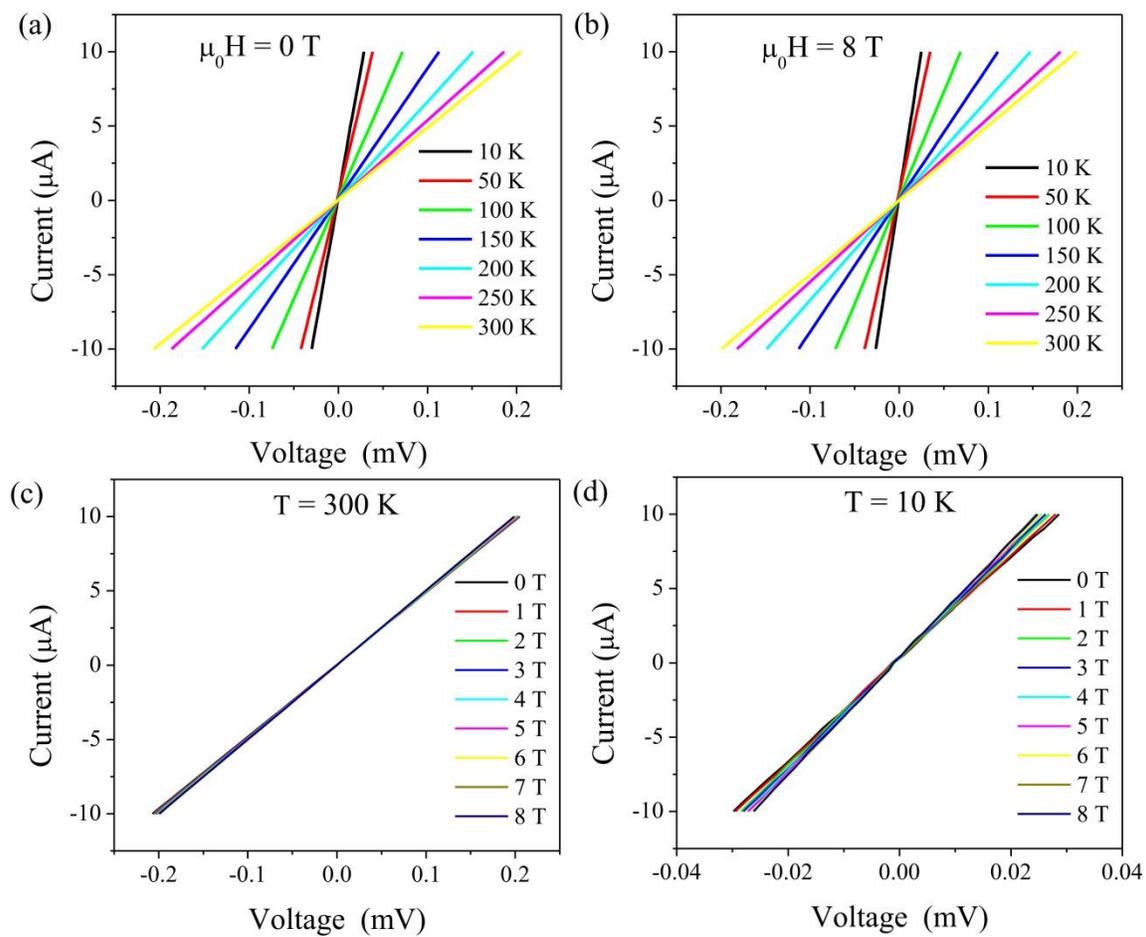

**Figure 5**

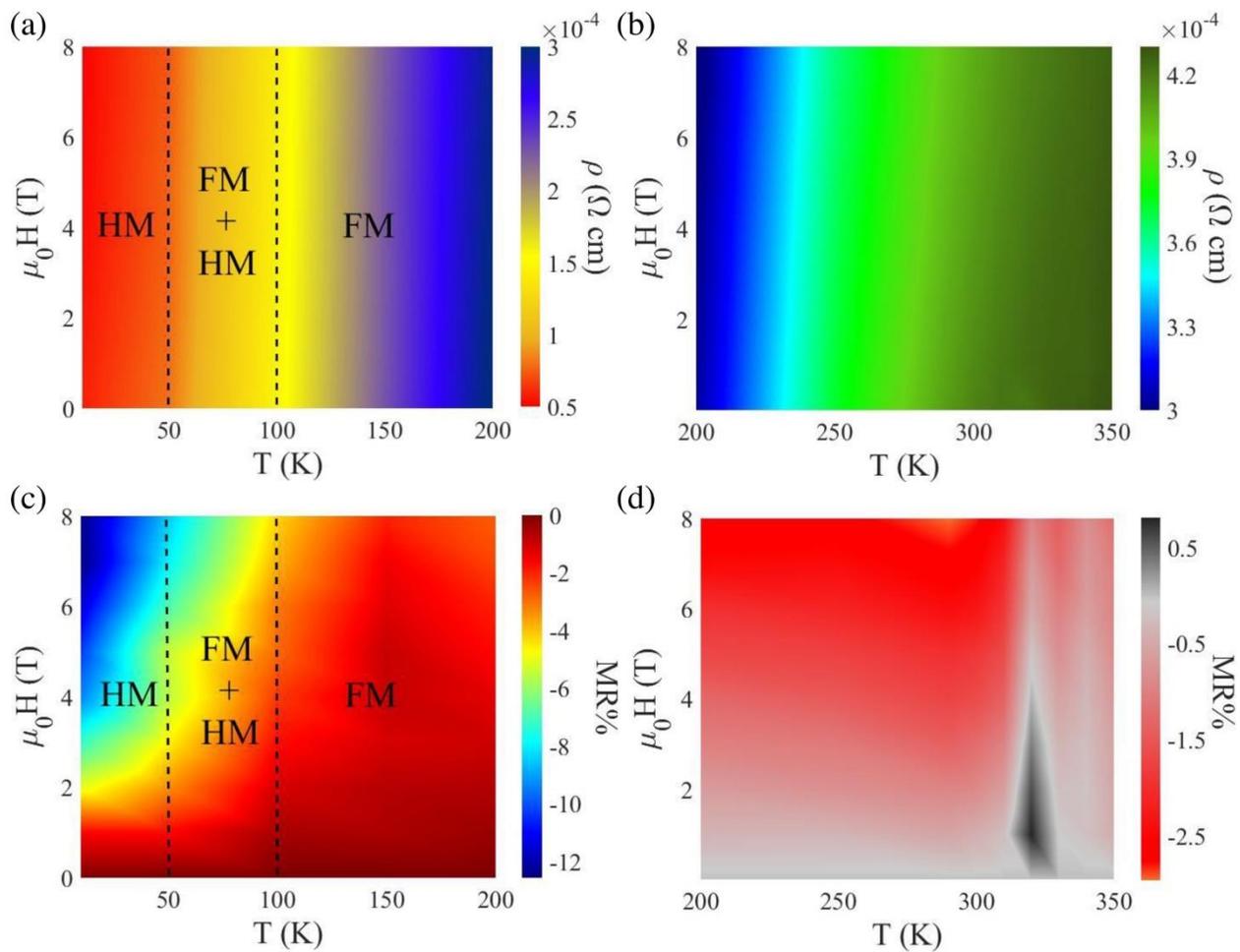